\documentclass[structabstract,fleqn]{aa}  
\usepackage{graphicx}
\usepackage{amssymb,amsmath}
\usepackage{subfig}
\usepackage{graphicx}
\usepackage{natbib}
\bibpunct{(}{)}{;}{a}{}{,} 
\usepackage{txfonts}
\usepackage{multirow}
\usepackage{hyperref}
\hypersetup{
    pdftoolbar=true,        
    pdfmenubar=true,        
    pdffitwindow=false,     
    pdfstartview={FitH},    
    pdftitle={The strongest gravitational lenses: III. The order statistics of the largest Einstein radii},    
    pdfauthor={Jean-Claude Waizmann},     
    pdfsubject={Draft},   
    pdfcreator={Jean-Claude Waizmann},   
    pdfproducer={Jean-Claude Waizmann}, 
    colorlinks=true,       
    linkcolor=blue,          
    citecolor=blue,        
    filecolor=blue,      
    urlcolor=blue           
}

\graphicspath{{./figures/}}

\usepackage{color}
\begin{document}

\title{The strongest gravitational lenses: III. The order statistics of the largest Einstein radii}

\titlerunning{The order statistics of the largest Einstein radii}


\author{J.-C. Waizmann \inst{1,2,3,4}\and M. Redlich \inst{5} \and M. Meneghetti \inst{3,2,4,6} \and M. Bartelmann \inst{5} }

\institute{Blue Yonder GmbH, Karlsruher Strasse 88, D- 76139 Karlsruhe, Germany\\ 
\email{jean-claude.waizmann@blue-yonder.com}
\and
Dipartimento di Fisica e Astronomia, Universit\`{a} di Bologna, viale Berti Pichat 6/2, I-40127 Bologna, Italy
\and
INAF - Osservatorio Astronomico di Bologna, via Ranzani 1, 40127 
Bologna, Italy
\and
INFN, Sezione di Bologna, viale Berti Pichat 6/2, 40127 Bologna, Italy
\and
Zentrum f\"ur Astronomie der Universit\"at Heidelberg, Institut f\"ur 
Theoretische Astrophysik, Albert-Ueberle-Str.~2, 69120 Heidelberg, Germany
\and
Jet Propulsion Laboratory, 4800 Oak Grove Drive, Pasadena, CA 91109, United States
}

\authorrunning{J.-C. Waizmann et al.}

\date{\emph{A\&A manuscript, version \today}}

\abstract
{The Einstein radius of a gravitational lens is a key characteristic. It encodes information about decisive quantities such as halo mass, concentration, triaxiality, and orientation with respect 
to the observer. Therefore, the largest Einstein radii can potentially be utilised to test the predictions of the $\Lambda$CDM model.}
{Hitherto, studies have focussed on the single largest observed Einstein radius. 
We extend those studies by employing order statistics to formulate exclusion 
criteria based on the $n$ largest Einstein radii and apply these criteria to the strong lensing 
analysis of 12 MACS clusters at $z>0.5$.}
{We obtain the order statistics of Einstein radii by a Monte Carlo approach, based on the 
semi-analytic modelling of the halo population on the past lightcone. After sampling the 
order statistics, we fit a general extreme value distribution to the first-order distribution,
which allows us to derive analytic relations for the order statistics of the Einstein radii.}
{We find that the Einstein radii of the 12 MACS clusters are not in conflict with the 
$\Lambda$CDM expectations. Our exclusion criteria indicate that, in order 
to exhibit tension with the concordance model, one would need to observe approximately 
twenty Einstein radii with $\theta_{\rm eff}\gtrsim 30\arcsec$, ten with 
$\theta_{\rm eff}\gtrsim 35\arcsec$, five with $\theta_{\rm eff}\gtrsim 42\arcsec$, or 
one with $\theta_{\rm eff}\gtrsim 74\arcsec$ in the redshift range $0.5\le z\le 1.0$ on 
the full sky (assuming a source redshift of $z_{\rm s}=2$). Furthermore, we find that, with 
increasing order, the haloes with the largest Einstein radii are on average 
less aligned along the line-of-sight and less triaxial. In general, the cumulative 
distribution functions steepen for higher orders, giving them better constraining 
power.}
{A framework that allows the individual and joint order distributions 
of the $n$-largest Einstein radii to be derived is presented. From a statistical point of view, 
we do not see any evidence of an \textit{Einstein ring problem} even for the largest Einstein 
radii of the studied MACS sample. This conclusion is consolidated by the large 
uncertainties that enter the lens modelling and to which the largest Einstein radii 
are particularly sensitive.}


\maketitle

\section{Introduction}\label{sec:intro}
The Einstein radius \citep{Einstein1936}, suitably generalised to non-circular lenses, is a key 
characteristic of every strong lensing system \citep[see e.g.][for recent reviews of gravitational 
lensing]{Bartelmann2010, Kneib2011, Meneghetti2013}. 
As a measure of the size of the tangential critical curve, it is very sensitive to a number 
of basic halo properties, such as the density profile, concentration, triaxiality, and the 
alignment of the halo with respect to the observer, but also to the lensing geometry, 
which is fixed by the redshifts of the lens and the sources \citep[see e.g.][]{Oguri2003, 
Oguri2009}. Moreover, the distribution of the sample of Einstein radii as a 
whole is strongly influenced by the underlying cosmological model. Here, not only 
cosmological parameters like the matter density, $\Omega_{\rm m}$, and the amplitude 
of the mass fluctuations, $\sigma_8$, but also the choice of the mass function, as well 
as the merger rate, have a strong impact. 

Recent studies gave rise to the so-called \textit{Einstein ring problem}, the claim that the 
largest observed Einstein radii \citep[see e.g.][]{Halkola2008, Umetsu&Broadhurst2008, 
Zitrin2011a, Zitrin2012} exceed the expectations of the standard 
$\Lambda$CDM cosmology \citep{Broadhurst&Barkana2008, Oguri2009, Meneghetti2011}. 
The comparison of theory and observations was performed by either comparing the largest 
observed Einstein radii with semi-analytic estimates of the occurrence probabilities of the 
strongest observed lens systems or with those found in numerical 
simulations. The most realistic treatment is certainly based on numerical simulations, which 
naturally include the impact of gas physics and mergers. However, for the statistical assessment 
of the strongest gravitational lenses, the number of simulation boxes and their sizes themselves 
are usually too small to sufficiently sample the extreme tail of the Einstein ring 
distribution. A sufficient sampling of the extreme value distribution of the largest Einstein radius 
roughly requires the simulation of $\sim 1000$ mock universes and a subsequent strong lensing analysis for each cluster sized halo \citep{Waizmann2012c}.

In this series of papers on the strongest gravitational lenses, we studied several 
aspects of the Einstein radius distribution. We utilised a semi-analytic approach that allows 
a sufficient sampling of the extreme tail of the Einstein radius distribution at the cost of a 
simplified lens modelling. In the first paper \citep[][herafter Paper I]{Redlich2012}, we 
introduced our method for the semi-analytic modelling of the Einstein radius distribution. We then 
studied the impact of cluster mergers on the optical depth for giant gravitational arcs 
of selected cluster samples and on the distribution of the largest Einstein radii. The 
second work \citep[][herafter Paper II]{Waizmann2012c} focussed on the extreme value 
distribution of the Einstein radii and the effects that strongly affect it, such as triaxiality, 
alignment, halo concentration, and the mass function. We could also show that the largest 
known observed Einstein radius at redshifts of $z>0.5$ of MACS J0717.5+3745 
\citep{Zitrin2009, Zitrin2011a, Medezinski2013} is consistent with the $\Lambda$CDM expectations. Now, in 
the third paper of this series, we extend the previous works by applying order statistics 
to the distribution of Einstein radii. Inference based on a single observation is difficult 
for it is a priori unknown whether the maximum is really drawn from the 
supposedly underlying distribution, or whether it is an event caused by a very peculiar situation 
that was statistically not accounted for. This is particularly important for strong lensing systems, 
which are heavily influenced by a number of different physical effects. It is therefore desirable to formulate 
$\Lambda$CDM exclusion criteria that are based on a number of observations instead of a single event. 
This goal can be accomplished by means of order statistics. We obtain the order statistic by 
Monte Carlo (MC) sampling of the hierarchy of the largest Einstein radii, using the semi-analytic 
method from Paper I, which is based on the work of \cite{Jing2002, Oguri2003}, and \cite{Oguri2009}. 
By fitting the generalised extreme value distribution to the first-order distribution, we derive 
analytic expressions for all order distributions and use them to formulate $\Lambda$CDM 
exclusion criteria. In the last part, we finally compare the theoretical distributions with the results 
of the strong lensing analysis of 12 clusters of the massive cluster survey 
\citep[MACS,][]{Ebeling2001, Ebeling2007} at redshifts of $z>0.5$ by \cite{Zitrin2011a}.

This paper is structured as follows. In \autoref{sec:order}, we 
introduce the mathematical prerequisites of order statistics, followed by a brief summary of 
the method for semi-analytically modelling the distribution of Einstein radii in 
\autoref{sec:samER}. Then, in \autoref{sec:prepConsid}, we present first results of the MC 
sampling of the order statistics. Afterwards, in \autoref{sec:dist_order}, we study the order 
statistical distributions and derive exclusion criteria based on the $n$ largest Einstein radii. 
This is followed by a comparison with the MACS sample in \autoref{sec:MACSsample} and an 
introduction of the joint two-order distributions in \autoref{sec:joint_distributions}. In \autoref{sec:conclusions}, we briefly summarize our main results and finally conclude.

For consistency with our previous studies, we adopt the \textit{Wilkinson Microwave Anisotropy Probe 7--year} (WMAP7) parameters $(\Omega_{\Lambda 0}, \Omega_{{\rm m}0}, \Omega_{{\rm b}0}, h, 
\sigma_8) = (0.727, 0.273, 0.0455, 0.704, 0.811)$ \citep{Komatsu2011} throughout this work.

\section{Order statistics}\label{sec:order}
In this section, we briefly summarise the mathematical prerequisites of order statistics that are needed for the remainder of this work. A more thorough treatment can be found in the excellent textbooks of \cite{Arnold1992} and \cite{David&Nagaraja2003} or, in a cosmological context, in \cite{Waizmann2012d}.

Suppose that $X_1,X_2,\dots ,X_n$ is a random sample of a continuous population with the 
probability density function (pdf) $f(x)$ and the corresponding cumulative distribution function 
(cdf) $F(x)$. Then, the order statistic is given by the random variates ordered by magnitude 
$X_{(1)}\le X_{(2)}\le \cdots \le X_{(n)}$, where $X_{(1)}$ is the smallest (minimum) and $X_{(n)}$ 
denotes the largest (maximum) variate in the sample. The pdf of $X_{(i)}\,(1\le i\le n)$ of the $i$-th order is 
then found to be given by
\begin{equation}\label{eq:f_order}
f_{(i)}(x)=\frac{n!}{(i-1)!(n-i)!}\left[F(x)\right]^{i-1}\left[1-F(x)\right]^{n-i}f(x)\, .
\end{equation}
Correspondingly, the cdf of the $i$-th order is given by
\begin{equation}\label{eq:F_order}
F_{(i)}(x)=\sum_{k=i}^n\binom{n}{k}\left[F(x)\right]^k\left[1-F(x)\right]^{n-k}\, .
\end{equation}
The distribution functions of the special cases of the lowest and the highest values are then readily 
found to be
\begin{equation}\label{eq:cdf_min}
F_{(1)}(x)=1-\left[1-F(x)\right]^n\end{equation}
and 
\begin{equation}\label{eq:cdf_max}
F_{(n)}(x)=\left[F(x)\right]^n\, .
\end{equation}
For large sample sizes, both $F_{(n)}(x)$ and $F_{(1)}(x)$ can be described by a 
member of the general extreme value (GEV) distribution \citep{Fisher1928,  Gnedenko1943} as used in Paper II of this series. In this case, the cdf is given by
\begin{equation}
 G(x) = \exp{\left\lbrace -\left[1+\gamma \left(\frac{x-\alpha}{\beta}\right)\right]^
 {-1/\gamma}\right\rbrace}\, ,
 \label{eq:cdf_gev}
\end{equation}
where $\alpha$, $\beta$, and $\gamma$ are respectively the location-, scale-, and shape-parameter, which can
either be obtained directly from the data or from an underlying model assumption (see 
\cite{Coles2001}, for instance).

The distribution functions of the single-order statistics can be generalised to $n$-dimensional joint 
distributions. In this work, we do not go beyond the two-order statistics for which the joint pdf  
$X_{(r)},X_{(s)}\;(1\le r < s \le n)$ for $x<y$ reads as
\begin{align}
f_{(r)(s)}(x,y)=&\frac{n!}{(r-1)!(s-r-1)!(n-s)!}\nonumber\\
 &\times\left[F(x)\right]^{r-1}\left[F(y)-F(x)\right]^{s-r-1}\left[1-F(y)\right]^{n-s}\nonumber\\
 &\times f(x)f(y)\, .
 \label{eq:joint_pdf_2d}
\end{align}
The joint cumulative distribution function can be derived by directly integrating the above pdf or
by a direct argument, and is given by

\begin{align}
F_{(r)(s)}(x,y)=& \sum_{j=s}^{n}\sum_{i=r}^{j}\frac{n!}{i!(j-i)!(n-j)!}\nonumber\\
                                                &\times\,\left[F(x)\right]^{i}\left[F(y)-F(x)\right]^{j-i}\left[1-F(y)\right]^{n-j}\, .
 \label{eq:joint_cdf_2d}
\end{align}
We refer the interested reader to Appendix A of \cite{Waizmann2012d} for more details on the implementation of the order statistics.
 
\section{Semi-analytic modelling of the distribution of Einstein radii}\label{sec:samER}
The adopted method for the semi-analytic modelling of the distribution of Einstein 
radii has already been thoroughly presented in two previous papers of this series 
\citep[see][and references therein]{Redlich2012, Waizmann2012c}. Thus, we only 
repeat the most important definitions and relations that are needed to follow this work.
\subsection{Defining the Einstein radius}\label{sec:definingER}
Historically, the Einstein radius has been defined for axially symmetric lenses. This assumption 
is untenable for realistic lenses, which exhibit irregular tangential critical curves. One 
generalised definition of the Einstein radius for arbitrarily shaped lenses is the 
effective Einstein radius, $\theta_{\rm eff}$, which is defined as 
\begin{equation}
\theta_{\rm eff}\equiv\sqrt{\frac{A}{\pi}}\;,
\end{equation}
where $A$ is the area enclosed by the critical curve. In the rest of this work, we 
only consider the effective Einstein radius.
\begin{figure*}
\centering
\includegraphics[width=0.3\linewidth]{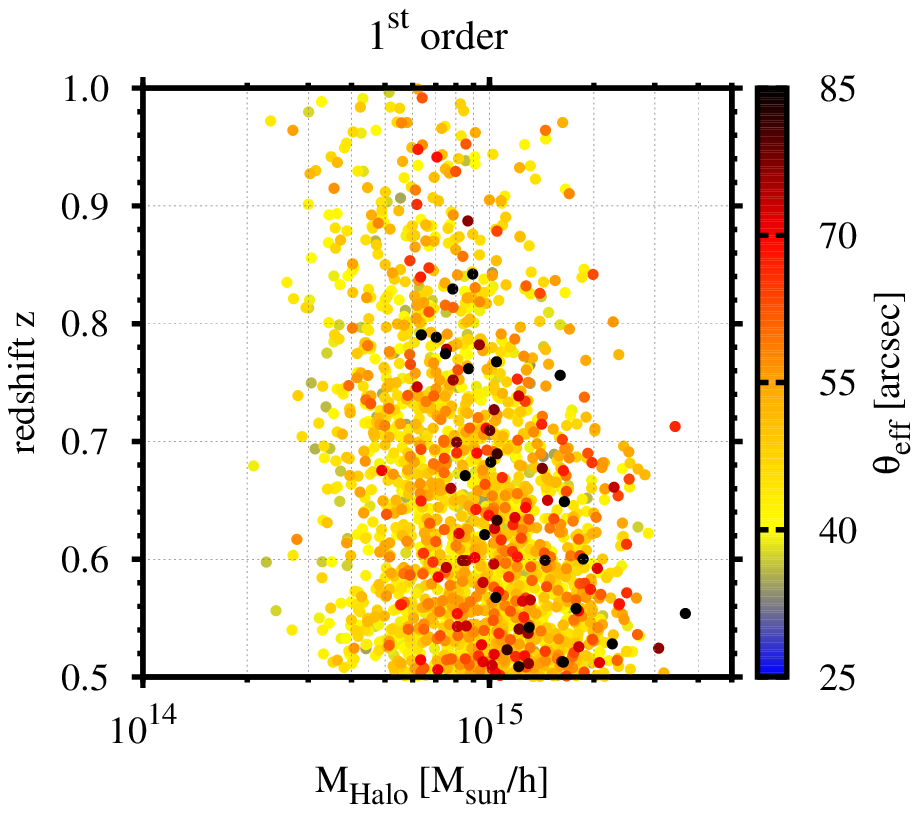}
\includegraphics[width=0.3\linewidth]{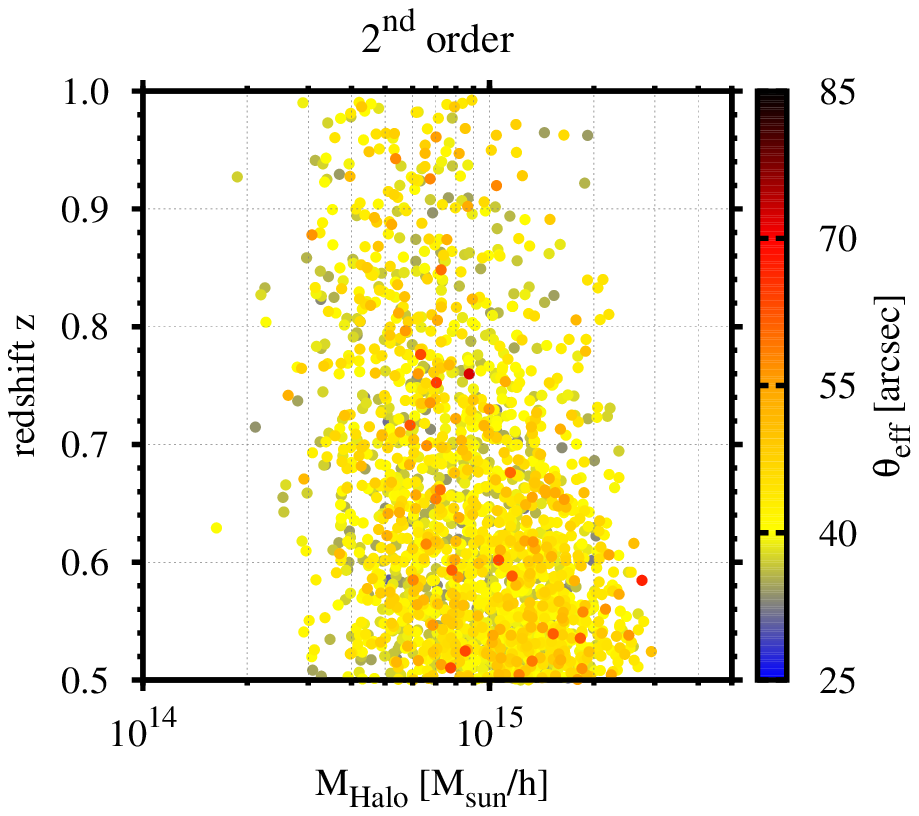}
\includegraphics[width=0.3\linewidth]{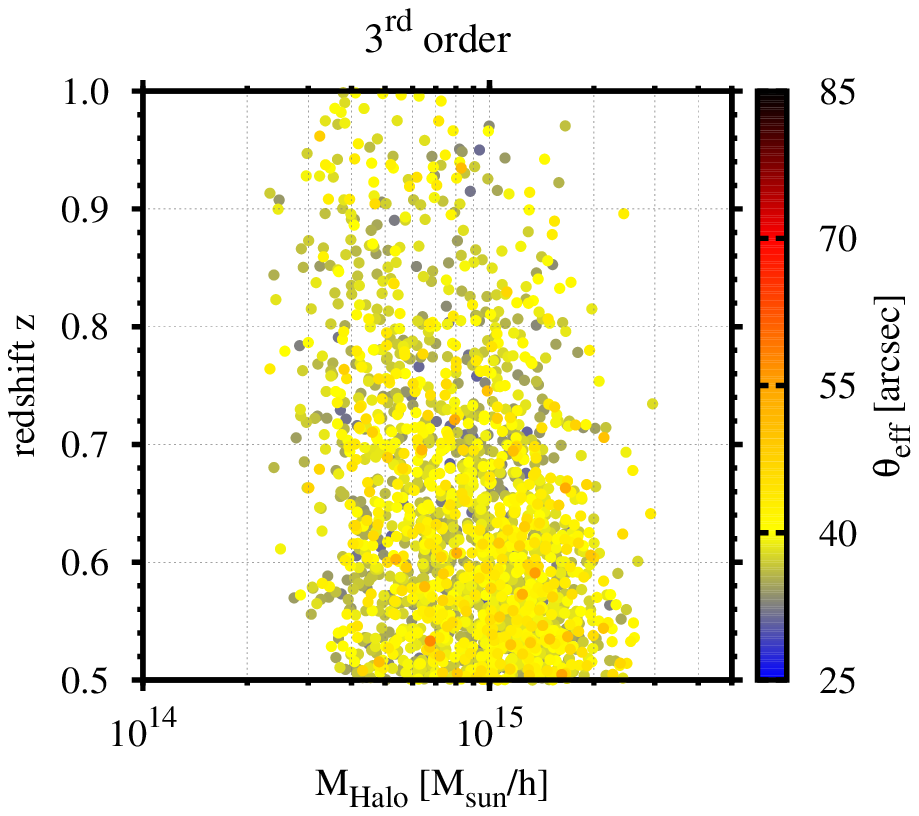}\\
\includegraphics[width=0.3\linewidth]{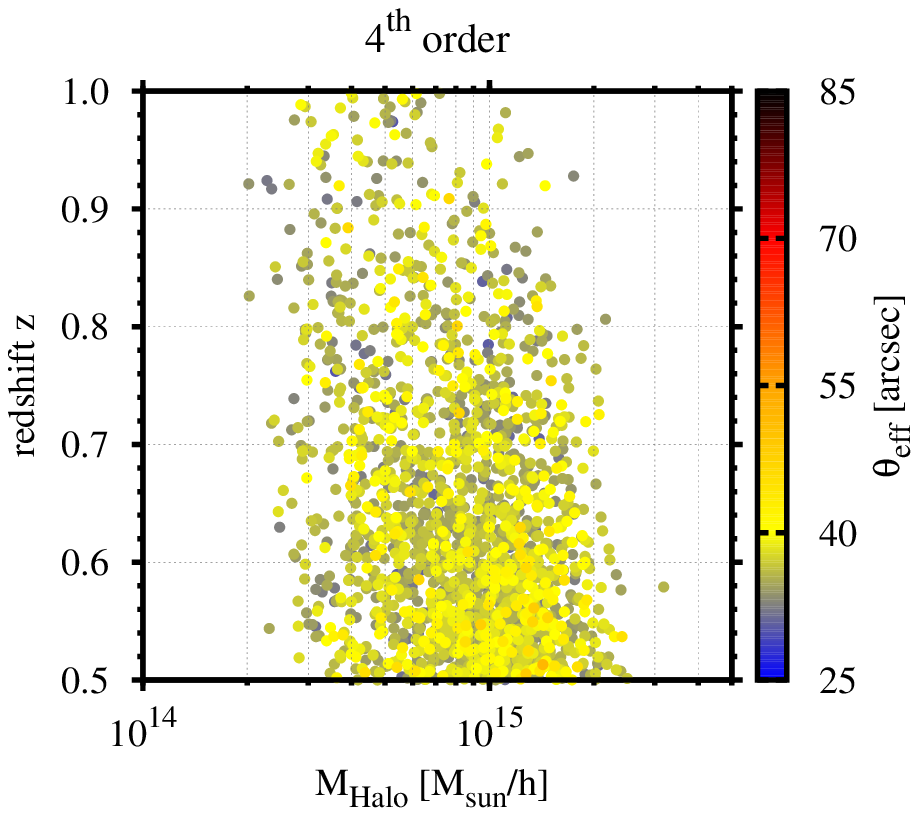}
\includegraphics[width=0.3\linewidth]{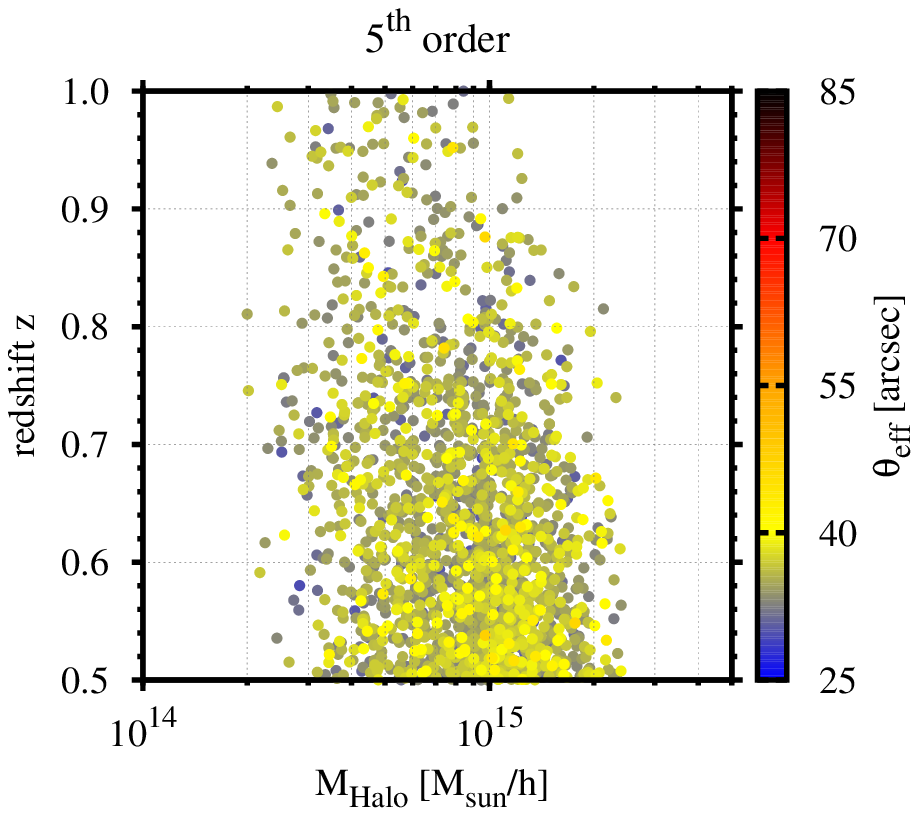}
\includegraphics[width=0.3\linewidth]{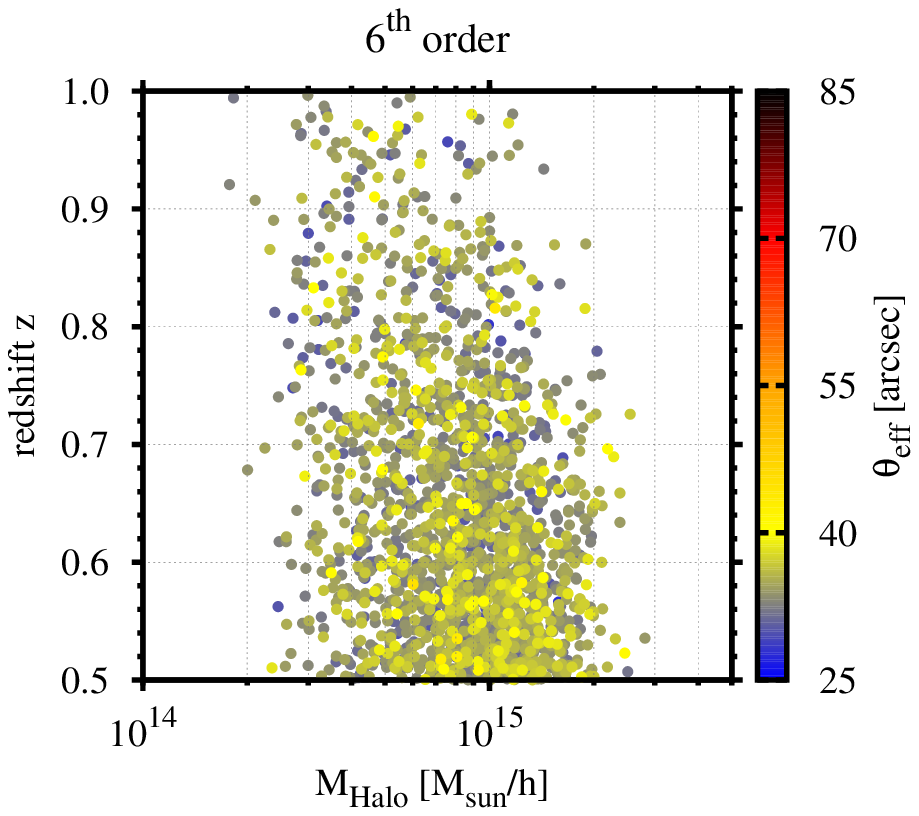}\\
\includegraphics[width=0.3\linewidth]{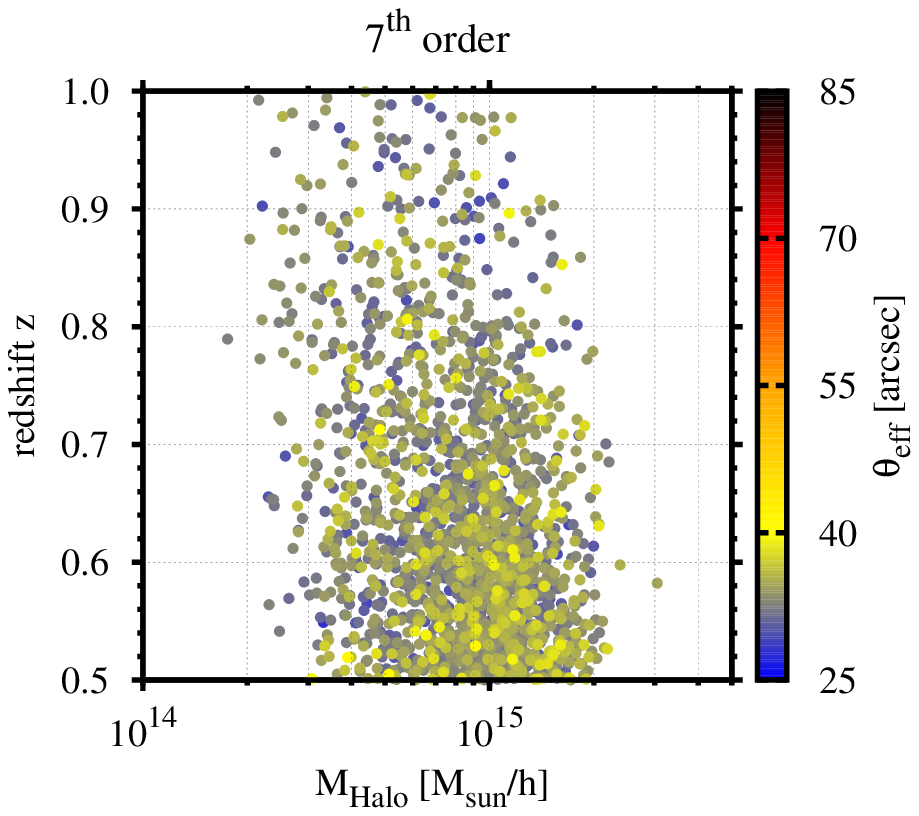}
\includegraphics[width=0.3\linewidth]{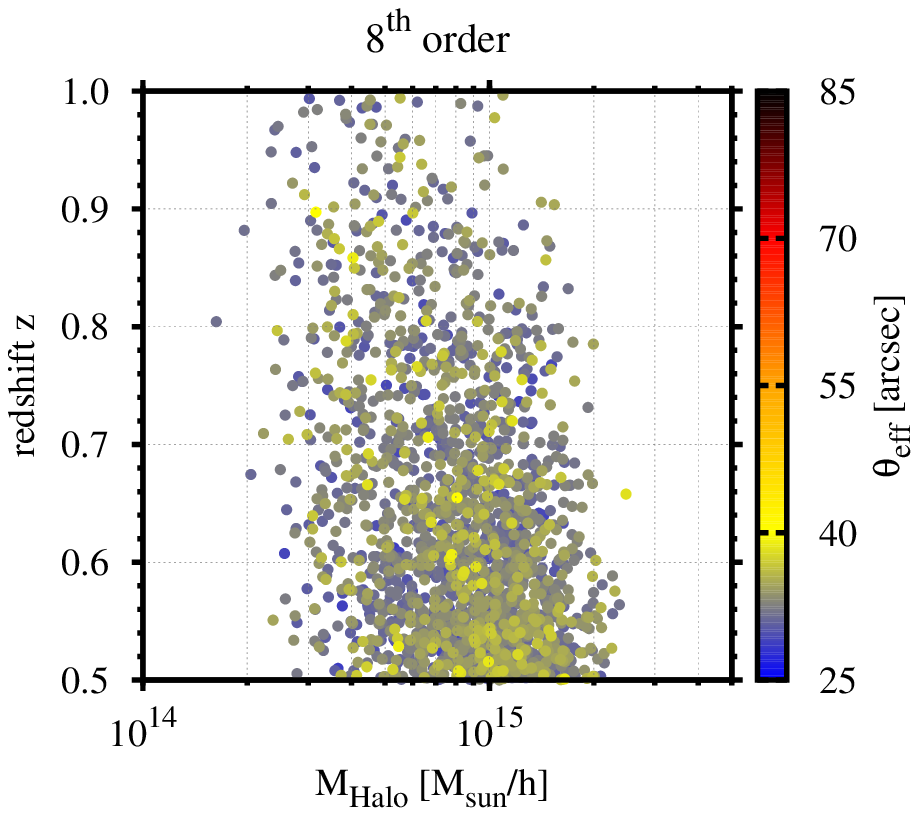}
\includegraphics[width=0.3\linewidth]{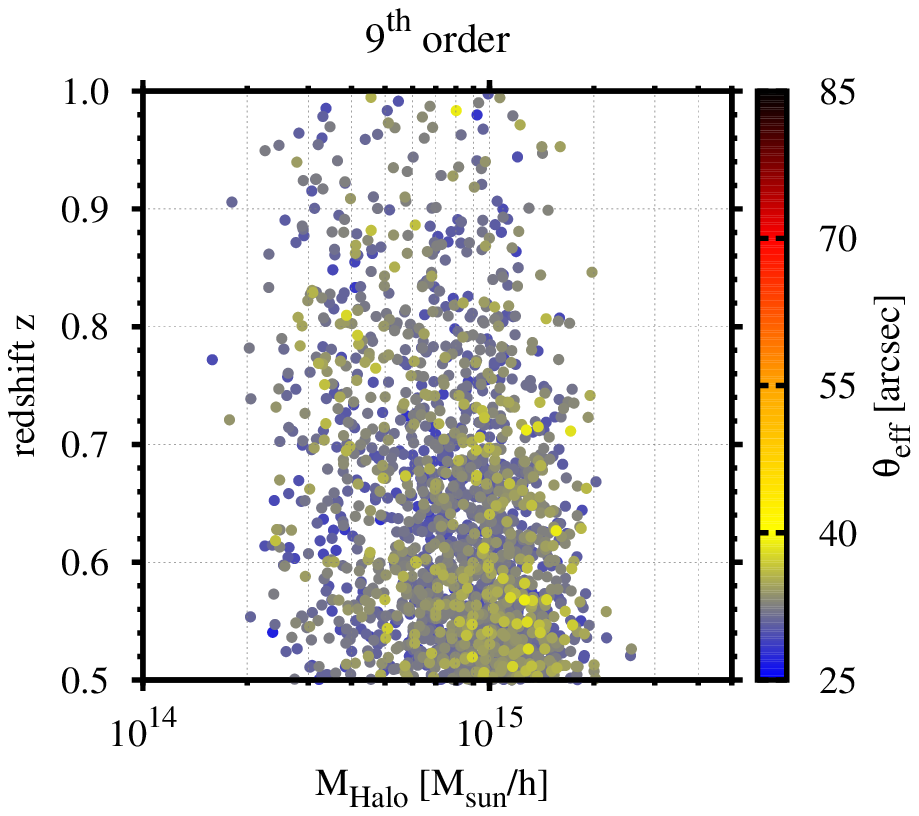}\\
\includegraphics[width=0.3\linewidth]{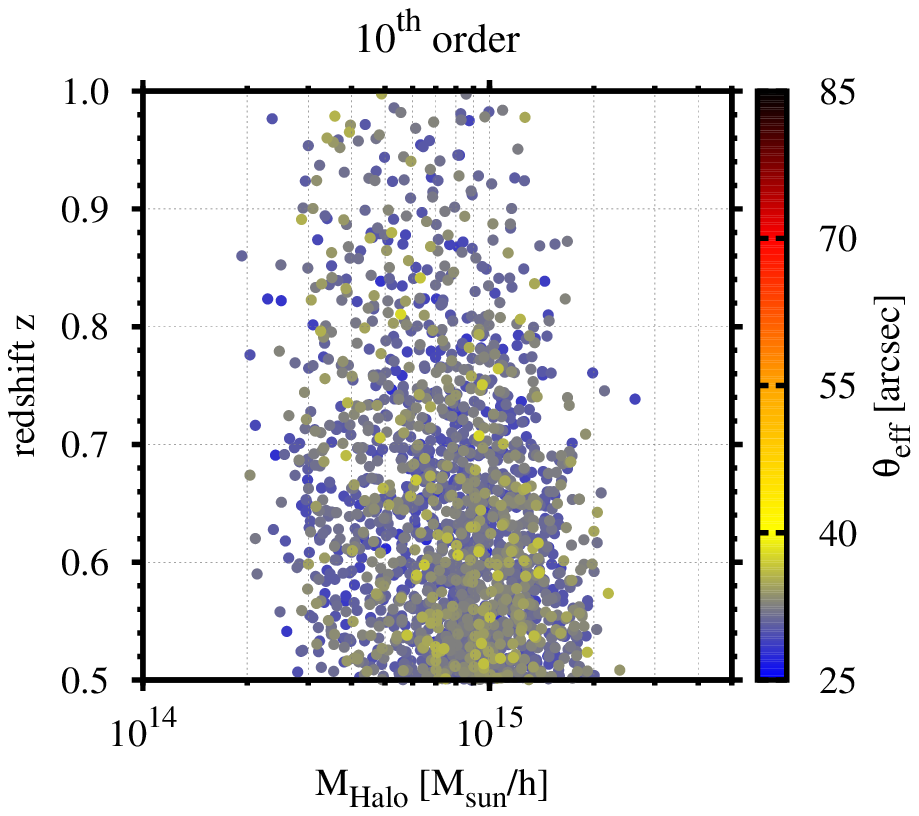}
\includegraphics[width=0.3\linewidth]{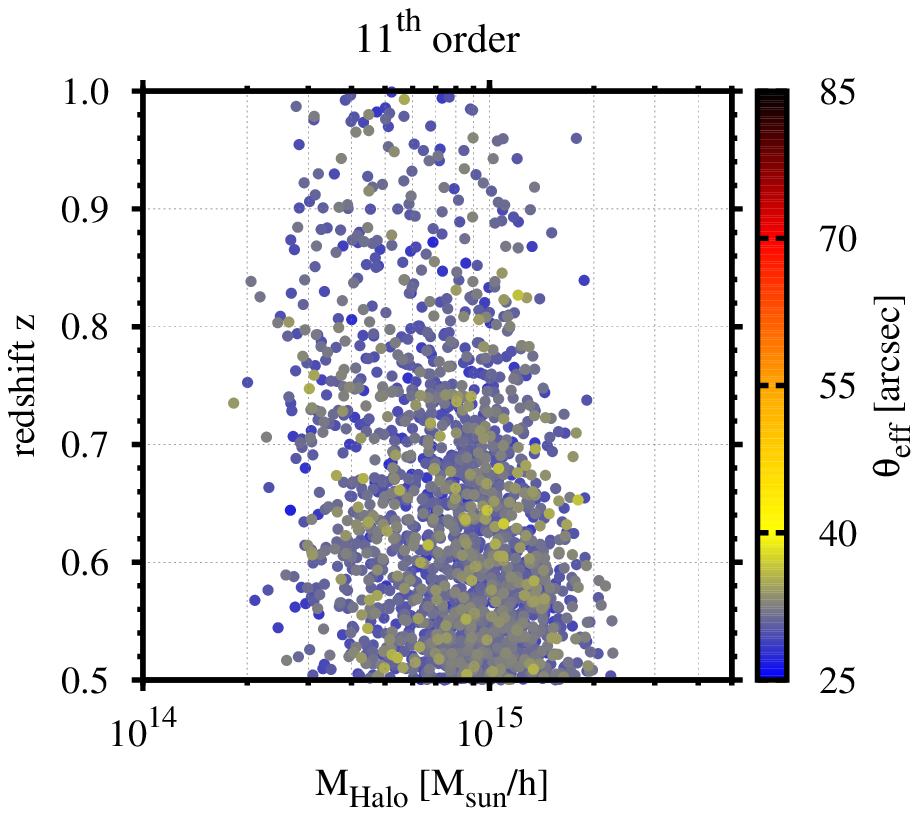}
\includegraphics[width=0.3\linewidth]{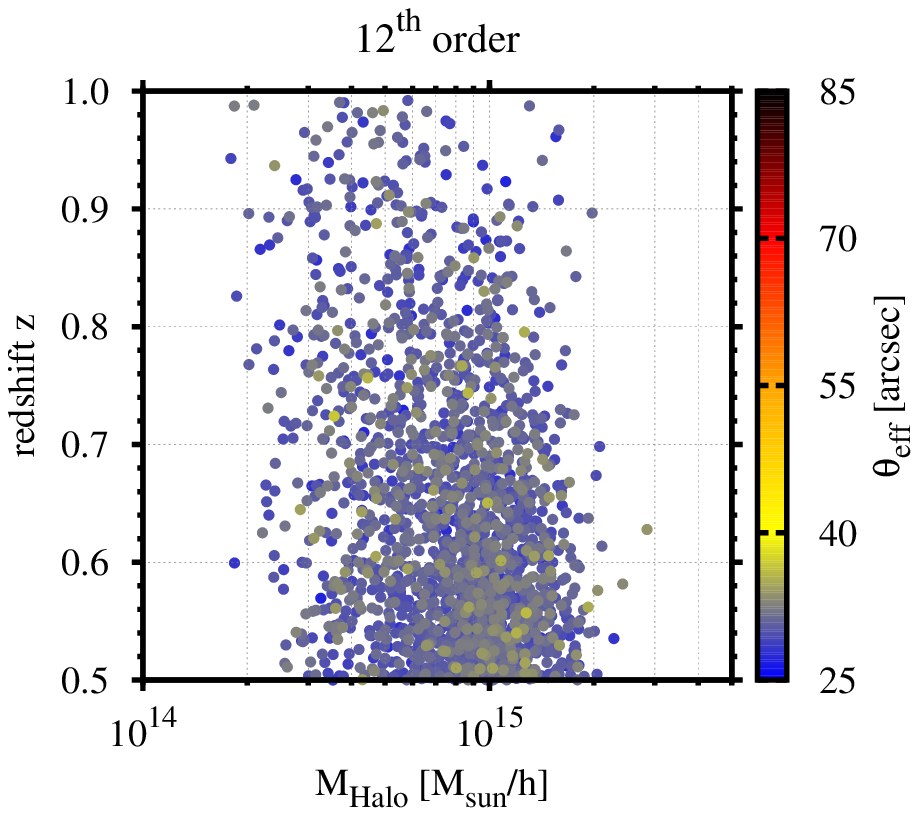}
\caption{Distribution in mass and redshift of $2\,000$ sampled values of the effective 
Einstein radius of the twelve largest orders as indicated above each panel. We assume 
the redshift interval of $0.5\le z\le 1.0 $ on the full sky, the \cite{Tinker2008} mass 
function, and a source redshift of $z_{\rm s}=2.0$. The colour encodes the size of the 
individual effective Einstein radii of a given order from each simulation run.}
\label{fig:Mz_distribution}
\end{figure*}
\subsection{Modelling triaxial haloes}\label{sec:singleHaloes}
Any realistic modelling of the distribution of Einstein radii has to account for 
halo triaxiality. An integral part of including triaxiality in the semi-
analytic modelling is the probability density functions of the axis ratios as they 
have been empirically derived by \cite{Jing2002}. Assuming the ordering 
$\left( a \leq b \leq c\right)$ of the axes, they read as\begin{align}
\label{eq:p_a}
p(a/c) &= \frac{1}{\sqrt{2\pi}\sigma_s} \exp\left[-\frac{(a_{\mathrm{sc}}-0.54)^2}
{2\sigma_{\mathrm{s}}^2}\right]\frac{\mathrm{d}a_{\mathrm{sc}}}{\mathrm{d}(a/c)} \; , \\ 
p(a/b|a/c) &= \dfrac{3}{2(1-r_{\mathrm{min}})}\left[1-\left(\dfrac{2a/b-1-r_{\mathrm{min}}}
{1-r_{\mathrm{min}}}\right)^2\right]  \; , \label{eq:p_ab}
\end{align}
where
\begin{equation}\label{eq:a_sc}
a_{\mathrm{sc}} = \frac{a}{c} \left( \frac{M}{M_*} \right)^{0.07[\Omega_{{\rm m}}(z)]^{0.7}} \; , 
\quad r_{\mathrm{min}} = \operatorname{max}\left(a/c,0.5\right) \; .
\end{equation}
Equation \eqref{eq:p_ab} holds for $a/b \ge r_{\mathrm{min}}$ and is zero otherwise. $M_*$ is the characteristic non-linear mass scale. For the width $\sigma_{\rm s} $ of the axis-ratio distribution $p(a/c)$, we adopt the value $\sigma_{\rm s} = 0.113$ as 
reported in \cite{Jing2002}.

In addition to the halo-triaxiality, the concentration parameter $c_{\rm e}$ plays an important 
role in defining a lensing system. The concentration is defined as $c_{\rm e}\equiv R_{\rm e}/R_0$,
where $R_{\rm e}$ is chosen such that the mean density within the ellipsoid of the major axis 
radius $R_{\rm e}$ is $\Delta_{\rm e}\Omega(z)\rho_{\rm crit}(z)$, with
\begin{equation}
\Delta_{\rm e}=5\Delta_{\rm vir}(z)\left(c^2/ab\right)^{0.75}.
\end{equation}
The virial overdensity, $\Delta_{\rm vir}(z)$, is approximated according to \cite{Nakamura1997}. 
For the pdf of the concentration parameter, \cite{Jing2002} find the log-normal distribution 
\begin{equation}
 p(c_{\rm e})=\frac{1}{\sqrt{2\pi}\sigma_{\rm c}} \exp\left[
-\frac{(\ln c_{\rm e}-\ln \bar{c}_{\rm e})^2}{2\sigma_{\rm c}^2}\right]\frac{1}{c_{\rm e}}, 
\label{eq:p_ce}
\end{equation}
with a dispersion of $\sigma_{\rm c} = 0.3$. The correlation between the axis ratio $a/c$ 
and the mean concentration can be modelled by the following relation \footnote{In Papers I and II, the exponent of (3/2) is missing due to a typographic error}  \citep{Oguri2003}
\begin{align}
\label{eq:ce}
\bar{c}_{\mathrm{e}} &= f_{\rm c} A_{\rm e} \sqrt{\frac{\Delta_{\mathrm{vir}}(z_{\rm c})}
{\Delta_{\mathrm{vir}}(z)}} \left( \frac{1 + z_{\rm c}}{1 + z} \right)^{3/2}\; , \\
f_{\rm c} &= \mathrm{max} \left\{0.3, 1.35 \exp \left[ - \left(\frac{0.3}{a_{\mathrm{sc}}} 
\right)^2 \right] \right\} \; ,
\label{eq:fc}
\end{align}
where $z_{\rm c}$ is the collapse redshift. The prefactor $f_{\rm c}$, defined in \autoref{eq:fc}, 
is a correction introduced by \cite{Oguri2009} in order to avoid unrealistically low 
concentrations for particularly small axis ratios $a_{\rm sc}$. Following \cite{Jing2002}, the 
free parameter $A_{\rm e}$ is set to a value of $A_{\rm e} = 1.1$. All expressions listed so 
far are valid for the standard $\Lambda$CDM model and an inner slope of the density profile 
of $\alpha_{\rm NFW} = 1.0$. As discussed in the previous works of this series, we force all 
sampled axis ratios $a_{\mathrm{sc}}$ to lie within the range of two standard deviations. By 
doing so, we avoid unrealistic scenarios with extremely small axis ratios and 
correspondingly low concentrations, where the lensing potential is dominated by
masses well outside the virial radius. 

A more detailed discussion of the semi-analytic modelling can be found in \cite{Jing2002}, 
\cite{Oguri2003}, and in the previous works of this series.
\begin{figure*}
\centering
\includegraphics[width=0.48\linewidth]{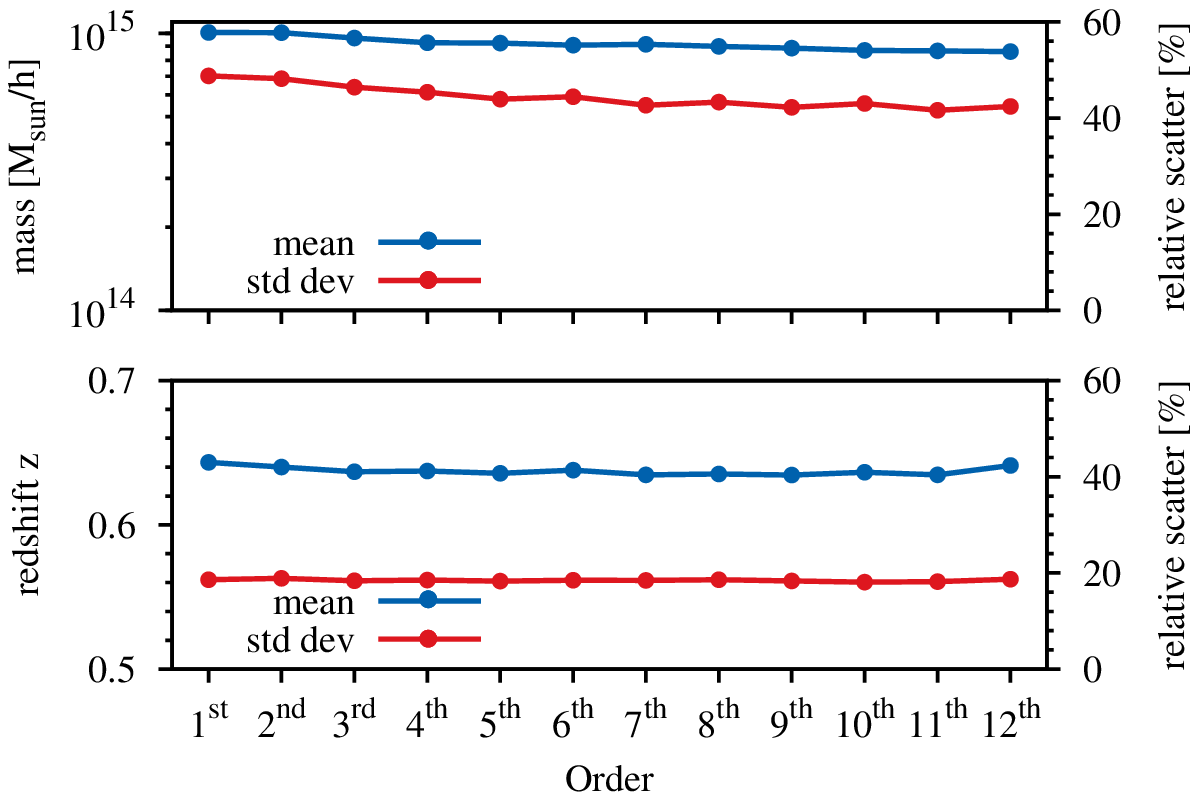}
\includegraphics[width=0.48\linewidth]{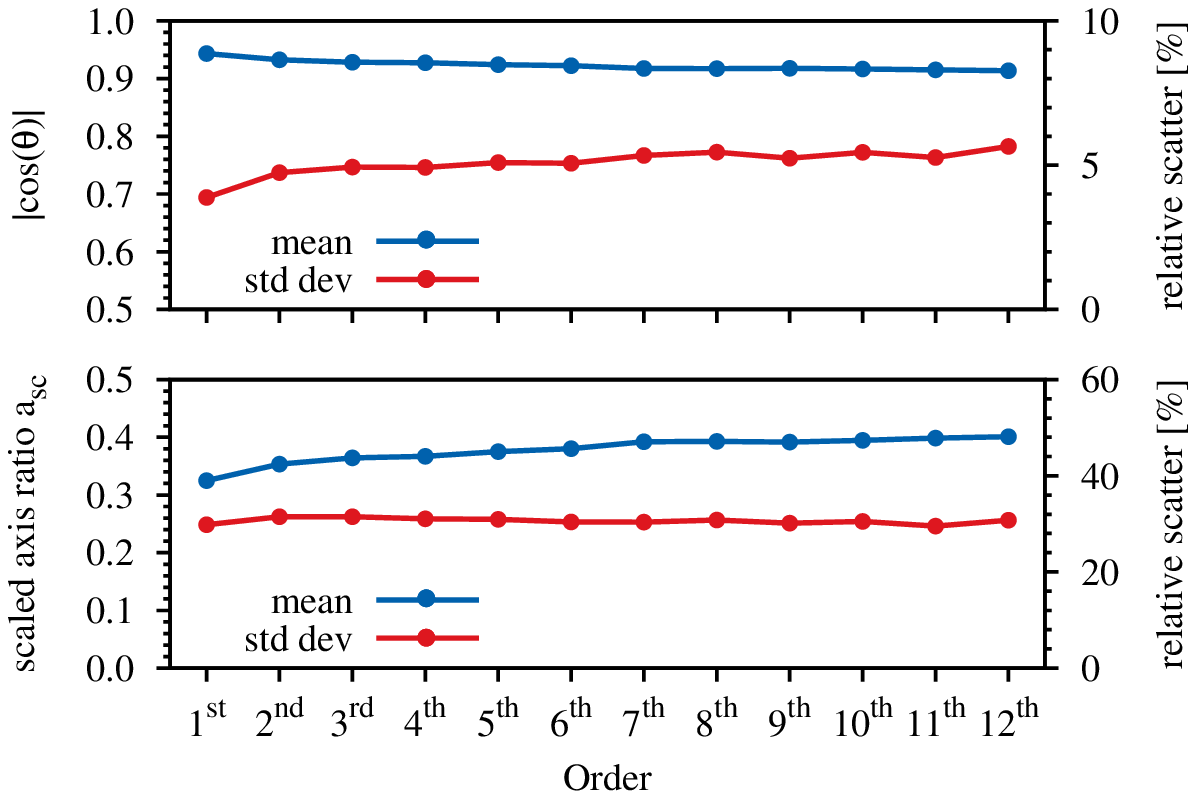}
\caption{Dependence of the sample mean (blue lines) and relative scatter (red lines) 
on the chosen order for the halo mass (upper left-hand panel), the redshift (lower left-hand 
panel), the alignment $|\cos(\theta)|$ (upper right-hand panel), and the scaled axis ratio, 
$a_{\rm sc}$ (lower right-hand panel). The red arrows denote constant shifts in the 
standard deviation in order to enhance the readability. The results are based on the samples 
presented in \autoref{fig:Mz_distribution}, comprising $2\,000$ sampled values in the 
redshift range of $0.5\le z\le 1.0 $ on the full sky.}
\label{fig:mean_mass_redshift}
\end{figure*}
\section{Results of the MC sampling of the order statistics}\label{sec:prepConsid}
The approach for sampling the order statistics of the Einstein radii is identical 
to the one used in Paper II. We create a large number of mock realisations of the cluster 
population on the past null cone, compute their strong lensing characteristics, and collect the Einstein 
radii of the largest orders. In this work, we use the \cite{Tinker2008} mass function for all calculations.
Because we eventually intend to compare our sampled distributions with the observed sample of $12$ MACS 
clusters in the interval $0.5\le z\le 0.6$, we only consider clusters in the redshift range $0.5\le z \le 1$
for our study of the order statistics of Einstein radii. This choice of the redshift interval also drastically
reduces the number of haloes that have to be simulated. To mimic the strong lensing analysis by 
\cite{Zitrin2011a}, we furthermore assume a fixed source redshift of $z_{\rm s}=2.0$. 

In Paper II, we showed that $~1000-2000$ realisations are sufficient for sampling the cdf of 
the largest Einstein radii and that all maxima stem from masses $M>2\times 10^{14}\,
M_\odot/h$. While the first statement will certainly hold for the order statistics, the second 
might not be valid any more for distributions of higher orders of the Einstein radius. To verify the second 
assumption, we decided to sample $2000$ mock realisations, adopting a lower mass limit of $M_{\rm lim}=10^{14}\,
M_\odot/h$. The distribution in mass and redshift for the first twelve orders is shown by
\autoref{fig:Mz_distribution}. It clearly demonstrates that, for the higher orders, only a few values
fall below the previous limit of $M>2\times 10^{14}\,M_\odot/h$. We will thus adopt the more conservative
lower mass limit of $M_{\rm lim}=10^{14}\,M_\odot/h$ for the rest of this work.

Furthermore, the distributions indicate that all orders stem from a wide range 
of masses. This tendency is confirmed by \autoref{fig:mean_mass_redshift}, which shows the dependence of the 
sample mean and the relative scatter in mass and redshift for the first twelve orders. 
It can be seen that the sample mean in mass (upper left-hand panel) weakly drops 
with increasing order and exhibits a large relative scatter of $\sim 40$ per cent. 
Although, on average, the sample of the largest Einstein radii stems from massive clusters with 
$M\sim 10^{15}\, M_\odot$, the large scatter indicates that it is not unlikely for notably less
massive clusters to contribute to this ordered list of very large Einstein radii. 
The sample mean of the clusters' redshifts (lower left-hand panel) is independent
of the order and shows a relative scatter of $\sim 20$ per cent. This is because
the distribution in redshift is mainly determined by the lensing geometry, which is obviously
independent of the considered order.

In addition to the mass and redshift distributions, it is interesting to examine how the orientation of the haloes and their 
scaled axis ratios depend on the different orders. For this purpose, we also calculated the sample mean and the relative scatter 
of the alignment, $|\cos(\theta)|$, and the scaled axis ratio, $a_{\rm sc}$, and present them in 
the right-hand panel of \autoref{fig:mean_mass_redshift}. A value of $|\cos(\theta)|=1$ 
corresponds to a perfect alignment of the halo's major axis with the line-of-sight of the observer. 
It can be seen (upper panel) that the mean alignment for the first twelve orders is high $(>0.9)$, but slightly decreases with increasing order. Thus, the higher the order, the more likely it happens that the haloes are no longer perfectly aligned with the observer's line-of-sight. This result can
be easily understood. For the very largest Einstein radii, all parameters (mass, orientation, concentration, etc.) simultaneously need to be most beneficial (in terms of strong lensing efficiency). For higher orders, a slightly disadvantageous setting of one parameter (e.g. a slightly misaligned halo) can still be compensated for by other halo properties. Nevertheless, the smallness of the relative scatter of the alignment with respect to the observer demonstrates that this property is an important characteristic of the sample of the largest Einstein radii. Closely related to the alignment is the elongation of the lensing-halo, which is encoded in the scaled axis ratio. In the lower right-hand panel of 
\autoref{fig:mean_mass_redshift}, we therefore present the dependence of the mean and relative scatter of $a_{\rm sc}$ on the order. A low value of $a_{\rm sc}$ indicates a very elongated system, while a halo with $a_{\rm sc}=1$ is spherical. The increase in the mean with order indicates that, with increasing order, it is more likely that the observed Einstein ring stems from a less elongated system. 

We summarise that, on average, the twelve largest Einstein radii stem from haloes with masses
$M\sim10^{15}\,M_\odot$. However, halo orientation and triaxiality (i.e. elongation) are influential factors that 
individually allow clusters with lower masses to produce very large Einstein radii. 
\begin{figure}
\centering
\includegraphics[width=0.95\linewidth]{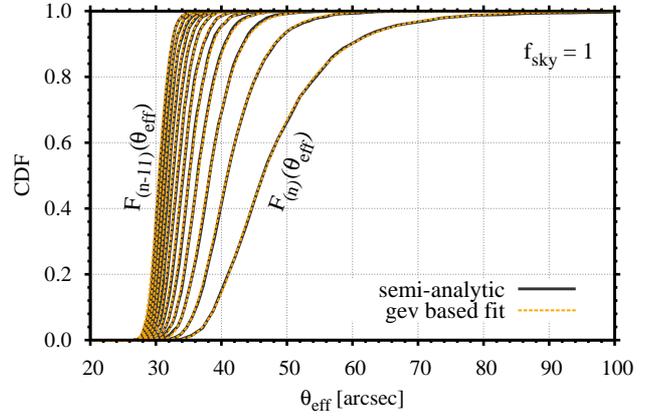}
\caption{Cumulative distribution functions of the first twelve order statistics of the 
effective Einstein radius. The black solid lines depict the distributions based on the 
semi-analytic MC sampling of $2\,000$ Einstein radii in the redshift range of 
$0.5\le z\le 1.0 $ on the full sky. The orange dashed lines denote fits of the distributions
of the maxima $F_{(n)}(\theta_{\rm eff})$ with a GEV distribution.}
\label{fig:Order_stat}
\end{figure}
\section{Comparison with observations}\label{sec:theory_vs_observations}
\subsection{The distributions of the order statistics of the effective Einstein radius}\label{sec:dist_order}
With a comparison to observed Einstein radii in mind, it is desirable to derive probability 
distributions of the order statistics of the effective Einstein radius. With the MC simulated data 
of the first twelve orders, presented in \autoref{fig:Mz_distribution}, we can now calculate the 
cdf of each order assuming full sky coverage, $f_{\rm sky}=1$. The result of this exercise is presented 
in \autoref{fig:Order_stat}, where the cdfs based on the MC data are presented. It can be seen nicely that the MC simulated data exhibits the steepening of the cdf with the 
increasing order that has also been found for the order statistics of the most massive and most 
distant galaxy clusters in \cite{Waizmann2012c}. As a result, while the first order $F_{(n)}$ is broad 
and allows the largest Einstein radius to be realised in a wide range, the higher orders are confined 
to an increasingly narrow range of radii. Therefore, the higher order cdfs can, in principle, be used 
to put tighter exclusion constraints based on the $n$-largest observed Einstein radii.

Because the distribution of the maxima, $F_{(n)}$, can be described by the GEV distribution from 
\autoref{eq:cdf_gev}, we fit this functional form to the semi-analytic cdf (best-fit parameters: $\alpha_{\rm eff}=43.52\pm 0.017$, $\beta_{\rm eff}=6.14\pm 0.026$ and $\gamma_{\rm eff}=0.13\pm 0.005$). Then, using the 
relation from \autoref{eq:cdf_max}, we can infer the underlying distribution, $F(\theta_{\rm eff})$, which can 
in turn be used to derive all order statistics. The result of this operation is shown in \autoref{fig:Order_stat}. It can be seen that the fitted order statistics 
match the semi-analytic distributions very well, confirming the consistency of the higher order cdfs.

\begin{figure}
\centering
\includegraphics[width=0.99\linewidth]{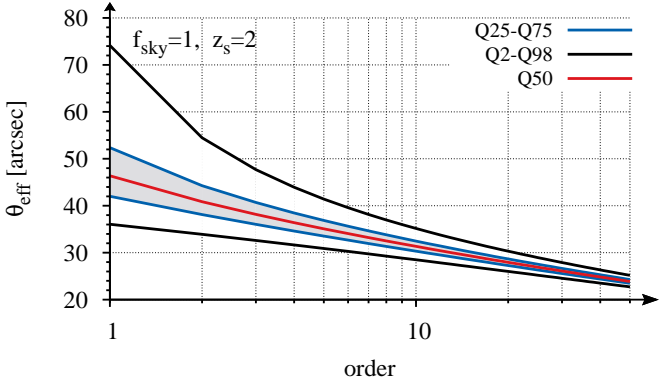}
\caption{Dependence of different percentiles of the effective Einstein radius on the order assuming 
full sky coverage and a source redshift of $z_{\rm s}=2$. The red line depicts the median (Q50), the blue 
lines the inner quartile range (Q25-Q75), and the black lines the 2-percentile (Q2,lower curve) and the 
98-percentile (Q98, upper curve). The latter can be used as $\Lambda$CDM exclusion criterion given 
the assumptions.}
\label{fig:exclusion_effER}
\end{figure}

The twelfth largest order statistic, $F_{(n-11)}(\theta_{\rm eff})$, (leftmost cdf in 
\autoref{fig:Order_stat}) indicates that one expects a dozen of Einstein radii with roughly $\theta_{\rm eff}
\gtrsim 30\arcsec$ on the full sky and assuming a source redshift of $z_{s}=2$. Applying order statistics 
to Einstein radii allows exclusion criteria to be formulated as a function of order, as presented in 
\autoref{fig:exclusion_effER}. Here, we show the dependence of different percentiles ($Q2$,$Q25$,$Q50$,
$Q75$, and $Q98$) on the order. Choosing the 98-percentile as $\Lambda$CDM exclusion criterion, 
one would need to find approximately twenty Einstein radii with $\theta_{\rm eff}\gtrsim 
30\arcsec$, ten with $\theta_{\rm eff}\gtrsim 35\arcsec$, five with $\theta_{\rm eff}\gtrsim 42\arcsec$, 
or one with $\theta_{\rm eff}\gtrsim 74\arcsec$ on the full sky, in order to claim tension with respect to the 
$\Lambda$CDM expectations. The presented exclusion criteria can be considered conservative because we modelled the distribution of Einstein radii using the simple semi-analytic method that does not incorporate cluster mergers. This choice was mainly motivated by the following reasons. Firstly and most importantly, in Paper II, we demonstrated that the precise choice of the mass function has a significant impact on the statistics of the largest Einstein radii. The \citet{Tinker2008} mass function is broadly considered to be more accurate than the original \citet{Press1974} mass function. However, it is a non-trivial task to self-consistently adapt merger tree algorithms to a given (empirical) mass function like the \citet{Tinker2008} mass function, among others. Secondly, we aimed to derive constraints that do not depend on the uncertainties (e.g. the simplified merger kinematics) of our semi-analytic method for modelling cluster mergers. Thirdly, the required computing time drops
significantly when cluster mergers are ignored. In particular, the first two points were important for our intention to minimise model uncertainties, thereby arriving at solid and rather conservative exclusion criteria. As discussed in Paper I, cluster mergers will certainly shift the cdfs to even higher values of $\theta_{\rm eff}$. We discuss the use of the order statistical distribution for falsification experiments in more detail 
in the following section.

For convenience, we use the fitted distributions for the rest of this work. Any small error 
introduced by this choice will be negligible in comparison to the unavoidable, still present model uncertainties
such as the precise shape of the mass function and the uncertainty in $\sigma_8$ (see Paper II, for a more detailed discussion).
\subsection{Comparison with the MACS sample}\label{sec:MACSsample}
We now intend to compare the theoretical order statistics with the effective Einstein radii that are based 
on the strong lensing analysis of a complete sample of twelve MACS clusters by \cite{Zitrin2011a} with 
$z>0.5$. Their parametric strong-lensing analysis \citep[see e.g.][for more details on the method]
{Broadhurst2005,Zitrin2009b,Zitrin2009c} of HST/ACS images of all twelve objects assumes a constant 
source redshift of $z_{\rm s}=2$, identical to the one used for deriving the order statistics. To 
compare the observed sample to the order statistical distributions, we sort all Einstein radii by size 
and list them in \autoref{tab:MACS_clusters}.

\begin{figure}
\centering
\includegraphics[width=0.95\linewidth]{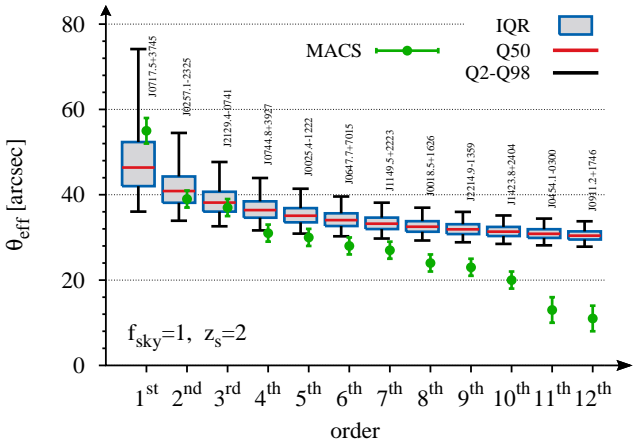}
\caption{Box-and-whisker diagram for the comparison of the order statistics with the 
twelve observed effective Einstein radii of the MACS cluster sample \cite{Zitrin2011a} as
listed in \autoref{tab:MACS_clusters}. For each order, the red lines denote the median 
($Q$50), the blue bordered grey boxes give the inner-quartile-range (IQR), and the black whiskers mark the 
range between the $2$ and $98-$percentile ($Q$2, $Q$98) of the theoretical distribution. 
The green error bars represent the observed effective Einstein radius.}
\label{fig:box_whisker_MACS}
\end{figure}

The results of the comparison are presented in \autoref{fig:box_whisker_MACS} in the form of a 
box-and-whisker diagram. Defining outliers with respect to the $\Lambda$CDM expectations 
as observations that exceed the $98-$percentile, none of the twelve observed Einstein radii falls 
outside the expectations for the full sky. That all observed Einstein radii with order larger than 
five fall below the $2-$percentile, with a much steeper slope, is a clear indication that the MACS sample 
is incomplete in terms of the largest Einstein radii that are expected to be found in the redshift range of 
$0.5\le z\le 1.0$. This is not surprising because we showed in the first part of this work that the sample of 
the largest Einstein radii stems from a wide range in mass and so a much larger observed sample is 
required to achieve completeness for the largest Einstein radii. In this sense, our conclusion is that the 
sample of effective Einstein radii of the studied MACS sample is consistent with the $\Lambda$CDM 
expectations. Furthermore, PLANCK results \citep{PlanckXVI2013} indicate higher values of  $\Omega_{{\rm m}0}$ and $\sigma_8$ in comparison to the WMAP7 ones that shift the probability distributions of the different orders to higher masses,  hence rendering the MACS sample even more likely to be found.

It would be interesting, though beyond the scope of this work, to extend this type of analysis to include the conditional order statistics of cluster samples that are selected given other observables. 

We should note that the nominal survey area of the MACS survey comprises only a fraction of the full sky 
$(A_{\rm s}=22 735\, {\rm deg}^2)$, which will shift the theoretical distributions to slightly lower values of 
$\theta_{\rm eff}$. The conclusion from above, however, should still hold, considering that we neglected
the impact of cluster mergers in our modelling (not all MACS clusters can considered to be relaxed) and, as discussed before, these events will substantially shift
the expectations to higher values of $\theta_{\rm eff}$.  Furthermore, a plethora of uncertainties that
enter the modelling of Einstein ring distributions will, from a statistical point of view, widen the range of the theoretical distributions. Those uncertainties comprise, for example, the uncertainty of the mass function at very high masses and the uncertainty in distribution of extreme halo axis ratios, which both strongly affect the distribution of the largest Einstein radii.
\begin{table}
\centering
\caption{Summary of the results of the strong lensing analysis of all 12 MACS 
clusters by \cite{Zitrin2011a} relevant for this work. The clusters are ordered by effective 
Einstein radius.}
\begin{tabular}{lcccc}
  \hline\hline
  MACS & $z$ & $\theta_{\rm eff}$\tablefootmark{a} & $\simeq r_{\rm eff}$ & Mass\tablefootmark{b}\\
   & &(arcsec) &(kpc)&($10^{14} M_{\odot}$)\\
  \hline
  J0717.5+3745  & 0.546 & $55\pm3$ & 353 & $7.40^{+0.50}_{-0.50}$ \\
  J0257.1-2325  & 0.505 & $39\pm2$ & 241 & $3.35^{+0.58}_{-0.10}$\\
  J2129.4-0741  & 0.589 & $37\pm2$ & 246 & $3.40^{+0.60}_{-0.30}$\\
  J0744.8+3927  & 0.698 & $31\pm2$ & 222 & $3.10^{+0.10}_{-0.10}$ \\
  J0025.4-1222  & 0.584 & $30\pm2$ & 199 & $2.42^{+0.10}_{-0.13}$\\
  J0647.7+7015  & 0.591 & $28\pm2$ & 187 & $2.07^{+0.10}_{-0.10}$ \\
  J1149.5+2223  & 0.544 & $27\pm3$ & 173 & $1.71^{+0.10}_{-0.20}$\\
  J0018.5+1626  & 0.545 & $24\pm2$ & 154 & $1.46^{+0.10}_{-0.10}$\\
  J2214.9-1359  & 0.503 & $23\pm2$ & 142 & $1.25^{+0.10}_{-0.10}$\\
  J1423.8+2404  & 0.543 & $20\pm2$ & 128 & $1.30^{+0.40}_{-0.40}$\\
  J0454.1-0300  & 0.538 & $13^{+3}_{-2}$& 83 & $0.41^{+0.03}_{-0.01}$\\      
  J0911.2+1746  & 0.505 & $11^{+3}_{-1}$& 68 & $0.28^{+0.02}_{-0.01}$\\
  \hline
\end{tabular}\label{tab:MACS_clusters}
\tablefoot{ Part of the data is based on the work of \cite{Ebeling2007}.\\ \tablefoottext{a}
{The estimation of the effective Einstein radius assumes a source redshift of $z_{\rm s}=2.0$. } 
\tablefoottext{b}{Mass enclosed within the critical curve in $10^{14}\,M_\odot$.}  }
\end{table}
\subsection{Joint distributions}\label{sec:joint_distributions}
\begin{figure*}
\centering
\includegraphics[width=0.99\linewidth]{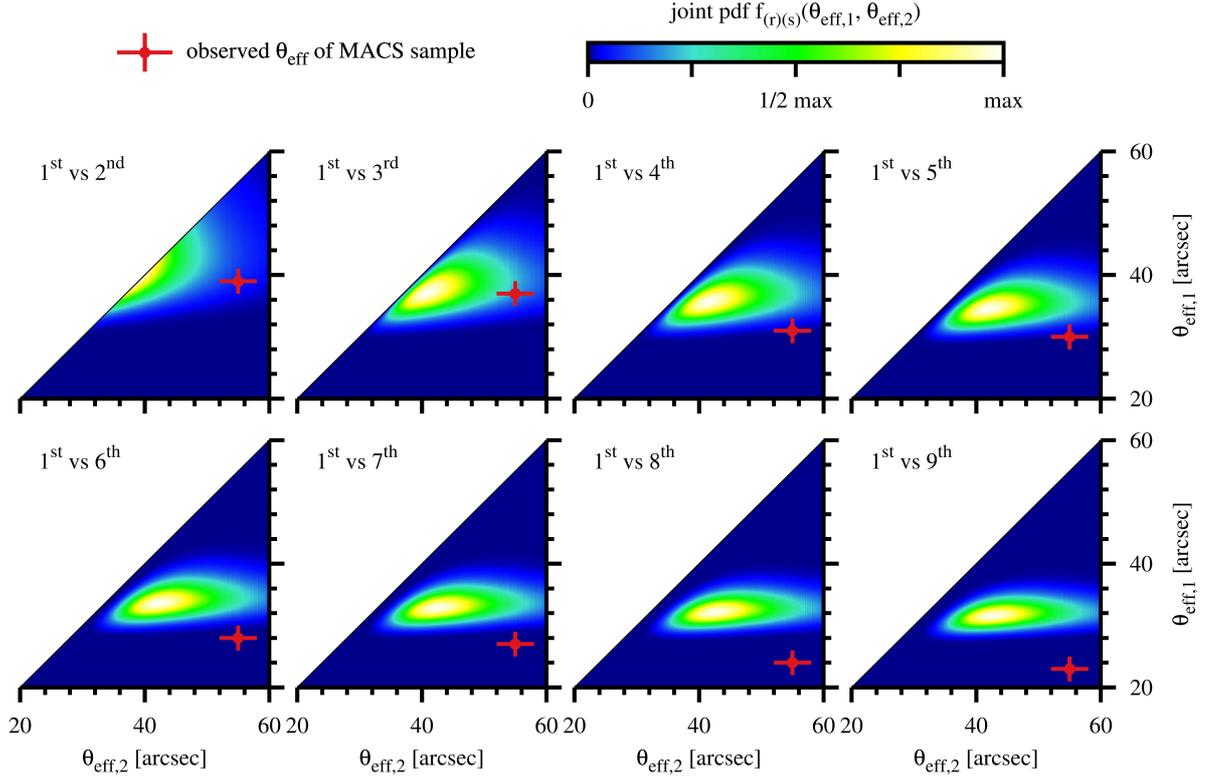}
\caption{Pdfs of the joint two-order statistics of the effective Einstein radius for different 
combinations of the first with higher orders as indicated in the upper right of each panel. 
The distributions are calculated for the redshift range of $0.5\le z\le 1.0 $ on the full sky. 
The color bar is set to range from $0$ to the maximum of the individual joint pdf in each 
panel. The red error bars denote the observed effective Einstein radii as listed in 
\autoref{tab:MACS_clusters}}
\label{fig:joint_distributions}
\end{figure*}

Apart from the study of the individual order statistics, it is also possible to derive joint 
distributions for different orders by means of \autoref{eq:joint_pdf_2d} and 
\autoref{eq:joint_cdf_2d}. We exemplarily present the joint pdfs for combinations 
of the first with the ninth largest orders in \autoref{fig:joint_distributions}. The pdfs are 
limited to a triangular domain due to the ordering constraint. It can be seen that, 
while for the combination with the second largest Einstein radius (upper left panel) we are more likely to find the values close to each other, for combinations with higher orders 
it is more likely that we find them to be realised with a larger gap. Furthermore, the peak of the joint 
pdf narrows for the smaller Einstein radius (y-axis) with increasing order. This is a direct 
result of the steepening of the cdf as shown in \autoref{fig:Order_stat}. 
In addition, we indicate the observed Einstein radii as the red error bars. That the 
red crosses fall with increasing order below the peak is a manifestation of the 
incompleteness of the sample as could also be seen for the individual order distributions 
in \autoref{fig:box_whisker_MACS}. In principle, the joint pdfs can also be extended to higher 
dimensions as outlined in \cite{Waizmann2012d} for the order statistics in mass and redshift. 

The joint pdfs shown in \autoref{fig:joint_distributions} also imply that the ratio of Einstein 
radii of different orders could itself be an important diagnostic. It may even be more robust 
than the Einstein radii themselves because perhaps the absolute calibration may drop out.  
\section{Summary and conclusions}\label{sec:conclusions}
In this work, a study of the order statistics of the largest effective Einstein 
radii has been presented. Using the semi-analytic method that we introduced in Paper I 
of this series, we sampled the distributions of the twelve largest Einstein radii in the 
redshift range of $0.5\le z\le 1.0$, assuming full coverage sky, the Tinker mass function, and 
a source redshift of $z_{\rm s}=2$. Thus, we generalise the statistical analysis of the 
single largest effective Einstein ring of Paper II to the one of the $n$-largest Einstein 
rings. Our main results can be summarised as follows. 
\begin{itemize}
\item The order statistics of the Einstein radii allows formulating $\Lambda$CDM 
exclusion criteria for the $n$-largest observed Einstein radii. We find that, in order 
to exhibit tension with the concordance model, one would need to observe roughly twenty Einstein radii with 
$\theta_{\rm eff}\gtrsim 30\arcsec$, ten with $\theta_{\rm eff}\gtrsim 35\arcsec$, 
five with $\theta_{\rm eff}\gtrsim 42\arcsec$, or one with $\theta_{\rm eff}\gtrsim 
74\arcsec$ in the redshift range $0.5\le z\le 1.0$, assuming full sky coverage and a 
fixed source redshift of $z_{\rm s}=2$. 
\item In the sample of semi-analytically simulated Einstein radii, the twelve largest radii 
stem from a wide range in mass. The sample mean in mass only slightly decreases with 
increasing order, while a large relative scatter of $\sim 40$ per cent is maintained. 
Additionally, we find that the haloes giving rise to the largest 
Einstein radii are on average well aligned along the line-of-sight and, with increasing rank, 
less triaxial. This finding supports the notion that, for the sample of the largest Einstein radii, 
triaxiality and halo alignment along the line-of-sight matter more than mass.  
\item For the sampled cdfs of the first twelve order statistics, we find a steepening of the 
cdfs with increasing order. This indicates that the higher orders are, in principle, more constraining. 
Using a GEV-based fit to the distribution of the maxima, we could show that the 
semi-analytic sample is self-consistent with the statistical expectation of the order statistics.
\item A comparison of the theoretically expected distributions with the MACS sample shows that the 
twelve reported Einstein radii of \cite{Zitrin2009} are consistent with the expectations 
of the concordance model. This conclusion would be consolidated further by (a) the inclusion 
of mergers and (b) the recent PLANCK cosmological parameters that indicate higher values of $\Omega_{{\rm m}0}$ and $\sigma_8$ in comparison to the WMAP7 parameters used in this analysis. Because we expect a large number of haloes to be potentially able to produce 
very large Einstein rings, the consistency of the studied MACS clusters with the $\Lambda$CDM 
expectations does not come as a surprise in view of the incompleteness of the sample.  
\item The method presented in this work allows calculation of joint distributions of an arbitrary 
combination of orders. As an example, we study the joint pdf of the two-order statistics and 
show that it is most likely that the largest and second largest Einstein rings are found to be 
realised with values very close to each other. For larger differences in the two orders, 
the gap between the observed values is expected to increase.
\end{itemize}
We presented a framework that allows the individual and joint order distributions 
of the $n$-largest Einstein radii to be derived. The presented method of formulating $\Lambda$CDM 
exclusion criteria by sampling the order statistical distribution is so general that it can easily 
be adapted to different survey areas and redshift ranges or can be included in an improved lens modelling. 
Such improvements will most certainly comprise the inclusion of mergers and realistic source 
distributions. From a statistical point of view, we do not see any evidence of an \textit{Einstein 
ring problem} for the studied MACS sample. This conclusion is consolidated by the large 
uncertainties that enter the modelling of the lens distribution, which the largest Einstein 
radii are particularly sensitive to.  
\begin{acknowledgements}
JCW acknowledges financial contributions from the contracts ASI-INAF I/023/05/0, 
ASI-INAF I/088/06/0, ASI I/016/07/0 COFIS, ASI Euclid-DUNE I/064/08/0, ASI-Uni 
Bologna-Astronomy Dept. Euclid-NIS I/039/10/0, and PRIN MIUR 2008 \textit{Dark 
energy and cosmology with large galaxy surveys}. MR thanks the Sydney Institute for 
Astronomy for the hospitality and the German Academic Exchange Service (DAAD) for 
the financial support. Furthermore, MR's work was supported in part by contract 
research \textit{Internationale Spitzenforschung II-1} of the Baden-W\"{u}rttemberg 
Stiftung. MB is supported in part by the Transregio-Sonderforschungsbereich TR 33 
\textit{The Dark Universe} of the German Science Foundation.
\end{acknowledgements}

\bibliographystyle{aa}

\end{document}